\documentclass[sigconf]{acmart}
\usepackage{graphicx}
\usepackage{amsmath}
\usepackage{booktabs}
\usepackage{algorithm}
\usepackage{algorithmic}
\usepackage{amsfonts}
\usepackage{multirow}
\usepackage{makecell}
\usepackage{subfigure}
\usepackage{color}
\usepackage{bm}
\usepackage{epstopdf}
\usepackage{url}
\usepackage[cal=cm]{mathalfa}
\usepackage{balance}
\usepackage{threeparttable}
\usepackage{lipsum}
\usepackage{enumitem}
\usepackage{pifont}

\newcommand{\argmin}{\operatornamewithlimits{argmin}}

\AtBeginDocument{%
  \providecommand\BibTeX{{%
    \normalfont B\kern-0.5em{\scshape i\kern-0.25em b}\kern-0.8em\TeX}}}

\begin{document}
\title{A Unified Framework for Cross-Domain Recommendation}
\renewcommand{\shorttitle}{UniCDR+}

 \author{Jiangxia Cao, Shen Wang, Gaode Chen, Rui Huang, Shuang Yang, Zhaojie Liu, Guorui Zhou}
\affiliation{
  \institution{Kuaishou Technology, Beijing, China}
  \country{\{caojiangxia, wangshen, chengaode, huangrui06, yangshuang08, zhaotianxing, zhouguorui\}@kuaishou.com}
 }

\begin{abstract}
In addressing the persistent challenges of data-sparsity and cold-start issues in domain-expert recommender systems, Cross-Domain Recommendation (CDR) emerges as a promising methodology.
 CDR aims at enhancing prediction performance in the target domain by leveraging interaction knowledge from related source domains, particularly through users or items that span across multiple domains (e.g., Short-Video and Living-Room).
For academic research purposes, there are a number of distinct aspects to guide CDR method designing, including the auxiliary domain number, domain-overlapped element, user-item interaction types, and downstream tasks.
With so many different CDR combination scenario settings, the proposed scenario-expert approaches are tailored to address a specific vertical CDR scenario, and often lack the capacity to adapt to multiple horizontal scenarios.
In an effect to coherently adapt to various scenarios, and drawing inspiration from the concept of domain-invariant transfer learning, we extend the former SOTA model UniCDR in five different aspects, named as UniCDR+.
Specifically, UniCDR+ is capable of modeling a range of characteristics, including static/sequential interactions, multi-hop neighbor signals, as well as hard and soft representation correlations. This flexibility allows UniCDR+ to adapt to several different CDR scenarios.
We conduct extensive experiments on 5 public CDR scenarios to show its effectiveness, and demonstrate its unique transfer ability across different interaction types and downstream tasks.
Besides, our work was successfully deployed on the Kuaishou Living-Room RecSys.

\end{abstract}


% \ccsdesc[500]{Computing methodologies~Neural networks}


\maketitle

\section{Introduction}

In industrial RecSys, following components are required: 
(1) On-device \textit{log collector} reports the raw users behaviours.
(2) \textit{Streaming data engine} assembles sample features from raw data.
(3) \textit{Offline platform} consumes streaming to optimize model.
(4) \textit{Online A/B test} combines several models/strategies to serving.
In a broad sense, the term \textbf{domain} refers to a set of samples adhering to the same distribution. For instance, applications such as  Kuaishou and TikTok offer a wide range of services to users, including areas such as Short-Video and Living-Room recommendations, as well as recommended E-shopping Content.
In this context, predictions related to Short-Videos and Living-Rooms can be viewed as two distinct domain data.

To accommodate diverse domain services, a fundamental approach involves deploying individual models specific to each service to consume corresponding data streams for training and updating. 
As a result, each model is exclusively trained on a single service's data stream, such as Short-video streaming data for one model, and Living-Room streaming data for another model.
However, in order to iterate models quickly, this route of deploying models separately for different services has become a basic setting in the industry.
Obviously, the above technical approach presents several challenges within the industrial context including:
Different domains may contain varying amounts of interaction data, e.g., the Short-Video domain data is more than 20x that of the Living-Room domain, this data imbalance makes the Living-Room RecSys could not fully learn the user interests, due to the small domain data sparsity  and cold-start problem (e.g., some user do not enter Living-Room page).
To this end, the Cross-Domain Recommendation (CDR) topic attracts a surge of industrial and academic attention to building more robust RecSys for data-scarce services.

\begin{figure}[t]
\begin{center}
\includegraphics[width=8cm,height=2.5cm]{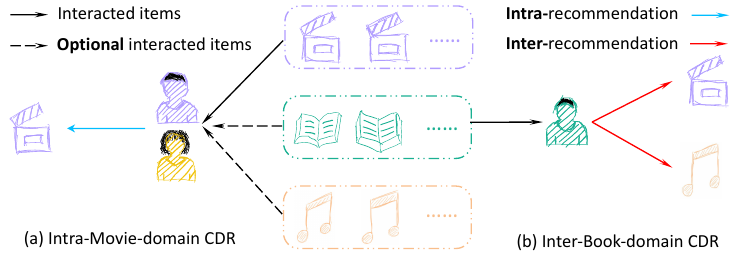}
\caption{Intra-/Inter- downstream tasks. Yellow user is overlapped user who has interaction in multiple domains, while purple/green users denote corresponding domain users.}
\label{downtask_task}
\end{center}
\vspace{-.4cm}
\end{figure}

To train a reliable CDR model, a consensus~\cite{crosssurvey} is that there exist some \textbf{overlapped elements} to bridge domains:
\begin{itemize}
	\item \textit{User-level overlap}: \textit{Short-Video} $\iff$ \textit{Living-Room}, viewed for a long time by the same user.
	\item \textit{Item-level overlap}: \textit{Online-shopping Short-Video} $\iff$ \textit{Online-shopping Living-Room}, featuring the same goods being sold.
\end{itemize}
Based on the overlapped elements, there are other factors that influence the \textit{training environment} and the method designing, such as the \textbf{domain numbers} (e.g., two~\cite{ddtcdr} or multiple~\cite{m3rec}) and \textbf{interaction type} (e.g., user-item static~\cite{disencdr} or sequential~\cite{pinet} interaction).
Aside from \textit{training scenario settings}, there are also have two \textbf{downstream tasks} at the \textit{model evaluation procedure}, for instance:
\begin{itemize}
	\item \textit{Intra-recommendation}~\cite{conet} for data-sparsity issue (in Fig.~\ref{downtask_task}(a)), which aims to fulfill users interest by other source domains, (Given Movie-domain user, predict next-basket Movies).	
	\item \textit{Inter-recommendation}~\cite{emcdr} for cold-start issue (in Fig.~\ref{downtask_task}(b)), which aims to transform the user interests from source to target domains, (Given Book-domain user, predict next-basket Movies/Songs).
\end{itemize}
With diverse training/test settings, many outstanding methods proposed~\cite{conet,emcdr}, but almost all of them are focused on one Scenario while lacking the competence to others.

\begin{figure}[t]
\begin{center}
\includegraphics[width=8cm,height=3.3cm]{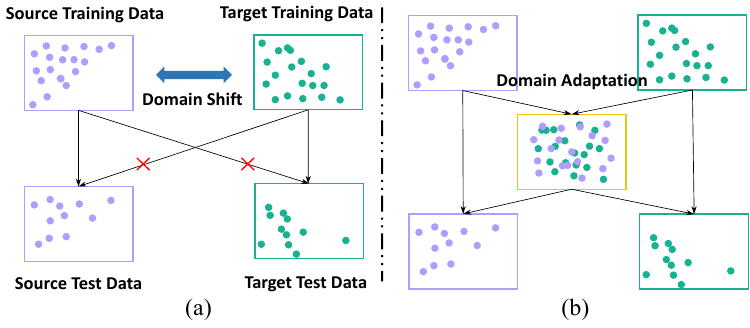}
\caption{Domain shift and adaptation of transfer learning.}
\label{transferlearning}
\end{center}
\vspace{-.4cm}
\end{figure}

The shattering phenomenon raises a question about whether can we find a way to serve various settings at the same time?
Our work aims to answer the question $\Rightarrow$ to build a unified framework to transfer knowledge across domains.
In particular, CDR can be regarded as an extension application of \textbf{transfer learning}, thereby CDR also follows the same assumptions in Fig.~\ref{transferlearning}(a).
That is, different domains can be seen as different data distributions, while the domain shift~\cite{transfer} impedes knowledge transferring.
Consequently, transfer learning aims at designing different adaptation strategies to alleviate the domain shift problem, e.g., learning a domain-invariant feature~\cite{cdrib} space to align distributions in Fig.~\ref{transferlearning}(b).

\begin{figure}[t]
\begin{center}
\includegraphics[width=8cm,height=3.3cm]{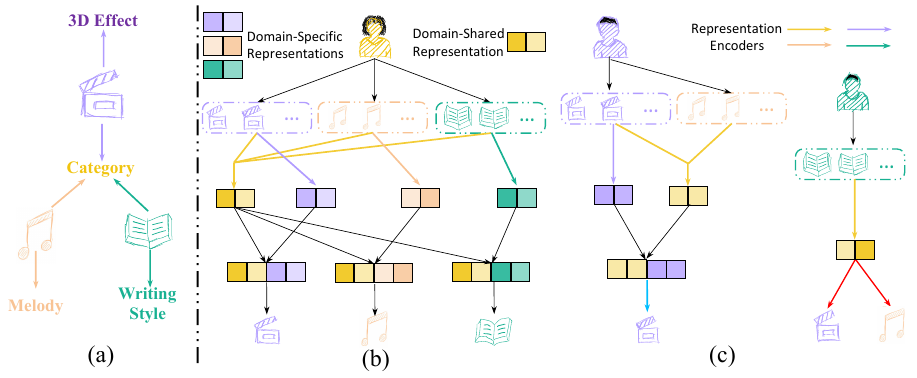}
\caption{A practical roadmap for unified CDR.}
\label{motivation}
\end{center}
\vspace{-.4cm}
\end{figure}

Inspired by the domain-invariant learning~\cite{invariant}, we find a similar phenomenon was shown in CDR (in Fig.~\ref{motivation}(a)).
This illustration shows three domains ``Movie'', ``Music'' and ``Book'', which reflect different information for user preference or item attributes, while some reliable information (i.e., ``\textit{Category}'') could provide positive effects for both domains and others preferences (i.e., ``\textit{2D/3D Effect}'') may lead to negative transferring.
Heuristically, the domain-specific preferences provide steady intra-domain interests, while shared preferences give \textbf{un-biased} intra-/inter- domain interests.

Therefore, a practical roadmap for various CDRs is: whatever the CDR training or evaluation setting is, the key insight is how to identify and assemble the domain- shared/specific information, \textbf{even different domains with different interaction types}.
Building on this, the pioneering UniCDR~\cite{unicdr} aims to transfer the domain-shared information across domains to serve different CDR scenarios.
As a typical work, UniCDR~\cite{unicdr} utilizes two stages:
\begin{itemize}
	\item \textit{Training stage}: for each user, learn to generate the domain-shared and specific representations (in Fig.~\ref{motivation}(b)).
	\item \textit{Evaluation stage}: predict intra-/inter- items by cooperating the shared/specific representations (in Fig.~\ref{motivation}(c)).
\end{itemize}
Nevertheless, UniCDR~\cite{unicdr} designing is focused on academic settings with static user-item interactions but neglects the multi-hop, sequential, and mixture interactions/features, which limits its capability in industrial RecSys.
In this work, we extend UniCDR~\cite{unicdr} to industrial fullrank~\cite{sim} procedure, and propose a more comprehensive and omnipotent framework, named UniCDR+.
Generally, we made several modifications on UniCDR in different aspects:
\begin{itemize}
	\item Considering great success of feature-enginnering and item-item multi-hop signals, we add corresponding components to fulfill the gap.
	\item To tackle different scale of item candidates in retrieval/fullrank process in RecSys chain, we provide several efficient interaction sequence aggregators.
	\item To capture invariant information more flexibly and multi-types actions, we introduce a soft objective to learn the domain-shared representations and a MMoE-style prediction module in our services.
\end{itemize}

Compared with academic methods, we extensively test UniCDR+ under 5 CDR scenarios to show its unique transferring ability with different branches SOTA works.
In our industrial setting, we conduct offline evaluation and online A/B test to demonstrate the effectiveness in Kuaishou Living-Room fullrank procedure, and UniCDR+ has now been deployed on the Kuaishou, serving million active users every day.
Overall, our contribution is that we extended the previous SOTA model UniCDR as UniCDR+, and performed analysis on public academic datasets and industrial services.

\section{Preliminary: CDR Scenarios Statement}

In this work, we ignore the user-item interaction explicit rating and suppose each user-item interaction as multiple binary implicit feedbacks, e.g., click, long-view and comment. 
Formally, we denote a domain data as $\mathcal{D}=(\mathcal{U}, \mathcal{F}_u, \mathcal{V}, \mathcal{F}_v, \mathcal{E}, \mathbf{A}$) , where the $\mathcal{U}/\mathcal{V}/\mathcal{E}$ are the user/item/interaction set respectively.
In industry, the $\mathcal{F}_u$/$\mathcal{F}_v$ are feature sets to describe user/item attributes, for example, including age, city, historical item interaction list for user, and cluster, multi-modal tags, empirical ratios for item.
The binary user-item bipartite graph adjacent matrix $\mathbf{A}\in\{0,1\}^{|\mathcal{U}|\times |\mathcal{V}|}$ is to describe the user-item edge from $\mathcal{E}$, where $\textbf{A}_{ij} = 1$ if user $u_i \in \mathcal{U}$ has observed interaction with item $v_j \in \mathcal{V}$ in $\mathcal{E}$ and $\textbf{A}_{ij} = 0$ otherwise.

Considering different training/evaluation environments, we assume there are existing three domains data $\{\mathcal{D}^X, \mathcal{D}^Y, \mathcal{D}^Z\}$, and categorize scenarios under the four aspects:
\begin{enumerate}
\item \textbf{Domain number}: two or multiple domains?
\item \textbf{Overlapped element}: bridged by users or items?
\item \textbf{Interaction type}: static or sequential interaction?
\item \textbf{Downstream task}: intra- or inter- recommendation?
\end{enumerate}
Considering the diverse combinations of the above aspects, our goal is to devise a unified framework to achieve the domain-shared information transferring idea, and we evaluate our framework under the 5 general academic CDR scenarios (in Table~\ref{CDR_scenarios}) and our service.

\begin{table}[t]
\footnotesize
\caption{The data format of five common CDR scenarios.}
% \resizebox{6.5cm}{
\setlength{\tabcolsep}{4pt}{
\begin{tabular}{c|cc|cc|cc|cc}
\toprule
\multirow{2}{*}{Scenarios} &\multicolumn{2}{c|}{\#Domain\_Num} &\multicolumn{2}{c|}{\#Over\_Elem} &\multicolumn{2}{c|}{\#Inter\_Type} &\multicolumn{2}{c}{\#Down\_Task} \\ \cline{2-3} \cline{4-5} \cline{6-7} \cline{8-9}
 &Dual &Multi &User &Item & Static & Seq & Intra & Inter   \\ 
\midrule
% \hline
Scenario 1 &\ding{52} &  & \ding{52} & &\ding{52}  &   &\ding{52} & \\
Scenario 2 &\ding{52} &  & \ding{52} & &\ding{52}  &  & &\ding{52}\\
Scenario 3 & & \ding{52} &  & \ding{52} &\ding{52}  &  &\ding{52} &\\
Scenario 4 &\ding{52} &  &\ding{52}  &  &  &\ding{52}  &\ding{52} &\\
Scenario 5 & & \ding{52} & \ding{52} & &\ding{52}  &  &\ding{52} &\\
\bottomrule
\end{tabular}
}
\label{CDR_scenarios}
\vspace{-.4cm}
\end{table}

\section{Preliminary: UniCDR Workflow}
\begin{figure}[t]
\includegraphics[width=8cm,height=3.7cm]{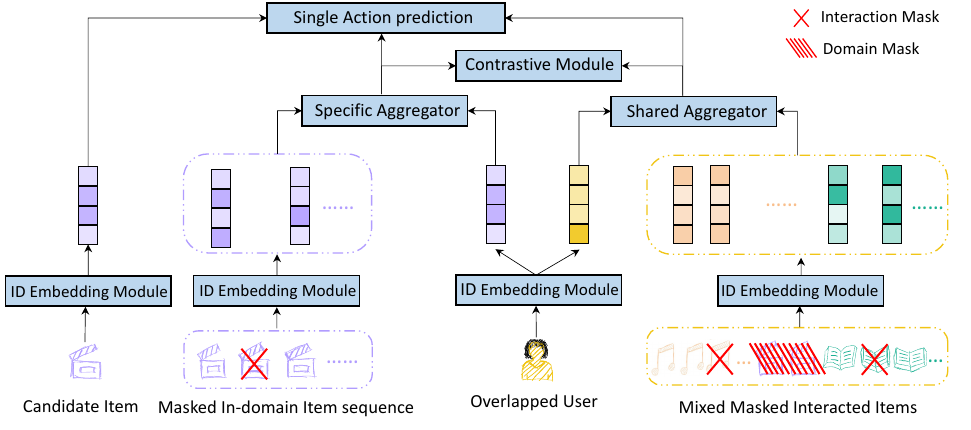}
\caption{The UniCDR framework.}
\label{originalunicdrframework}
\vspace{-.4cm}
\end{figure}

In this section, we first retrospect the former SOTA unified CDR method components, UniCDR~\cite{unicdr}.
Fig.~\ref{originalunicdrframework} presents training workflow of UniCDR under the Scenario 5, with a bird's-eye view, which follows a "two-tower" designing style to obtain the users domain-shared\&specific representations.

\subsection{Embedding Module}
%  In the embedding index look-up process, we embed 
\subsubsection{User/Item ID Embedding}
In UniCDR~\cite{unicdr}, the learning centroid is to recognize domain-shared information.
To consider such information easily, UniCDR introduces two types of embeddings to embed users and items for domain $X$, $Y$ and $Z$, i.e., the domain-specific initialization matrices $\mathbf{U}^{X}\in \mathbb{R}^{|\mathcal{U}^X|\times d}, \mathbf{U}^{Y} \in \mathbb{R}^{|\mathcal{U}^Y|\times d}$, $\mathbf{U}^{Z} \in \mathbb{R}^{|\mathcal{U}^Z|\times d}$ and domain-shared initialization matrix $\mathbf{U}^{S} \in \mathbb{R}^{|\mathcal{U}^X\cup\mathcal{U}^Y\cup\mathcal{U}^Z|\times d}$, where $d$ is the embedding dimension.  
Since the item sets are distinct in Scenario 5, UniCDR only use the domain-specific initialization matrices $\mathbf{V}^{X}\in \mathbb{R}^{|\mathcal{V}^X|\times d}$, $\mathbf{V}^{Y}\in \mathbb{R}^{|\mathcal{V}^Y|\times d}$, $\mathbf{V}^{Z}\in \mathbb{R}^{|\mathcal{V}^Z|\times d}$.

\subsection{Aggregators}
\label{seqagg}
The user interests extracting module is the most researched point in RecSys, UniCDR~\cite{unicdr} introduces three simple interaction aggregators from different aspects.

\subsubsection{Aggregators details}
For simplicity, we neglect the domain remarks $\{X, Y, Z\}$, and use embedding $\mathbf{u}$/$\{\mathbf{v}_1,\mathbf{v}_2,\mathbf{v}_3,\dots\}$ to represent user $u$'s behaviours $\mathcal{H}_u=\{v_1,v_2,v_3,\dots\}$.

% \subsubsection{Mean-pooling aggregator}
\noindent\textbf{(1) Mean-pooling aggregator}
As used in many early methods~\cite{item2vec}, the mean-pooling is efficiently to estimate user static interests:
\begin{equation}
% \small
 \footnotesize
	\begin{split}
	\mathbf{h} = \textsc{Mean}(\{\mathbf{v}_1,\mathbf{v}_2,\mathbf{v}_3,\dots\})\mathbf{W}_{\texttt{agg}},
	\end{split}
\label{meanpooling}
\end{equation}
where the $\mathbf{W}_{\texttt{agg}}\in \mathbb{R}^{d\times d}, \mathbf{h}$ denote the weight matrix and user's interests representation, respectively. 

\noindent\textbf{(2) User-attention-pooling aggregator}
Compared to the mean-pooling aggregator, the user-attention-pooling~\cite{attention} has the advantage of auto-detecting the personalized important items for users.
\begin{equation}
 \footnotesize
	\begin{split}
	\mathbf{h} = \textsc{Attention}(\mathbf{u}, \{\mathbf{v}_1,\mathbf{v}_2,\mathbf{v}_3,\dots\}),
	\end{split}
\label{userattention}
\end{equation}

\noindent\textbf{(3) Item-similarity-pooling aggregator}
Instead of user-attention, the item relationship is also an important perspective, which provides valuable clues to find implicit conjoint interests to generate personalized representation.
Thus, UniCDR introduces the item-item linear model EASE$^R$~\cite{ease} to generate item similarity relationship matrix $\mathbf{B}\in \mathbb{R}^{|\mathcal{V}|\times|\mathcal{V}|}$ by optimizing:
\begin{equation}
 \footnotesize
  	\argmin_{\mathbf{B}} \|\mathbf{A} - \mathbf{A}\mathbf{B}\|_F^2 + \lambda_F\|\mathbf{B}\|_F^2
\label{EASE}
\end{equation}
where $\lambda_F$ is a penalty hyperparameter.
Particularly, the Eq.(~\ref{EASE}) has the closed-form solution to pre-process the $\mathbf{B}$ to support our training (See~\cite{ease} for the detailed derivation).
Next, UniCDR can exploit the similarity $\mathbf{B}$ to obtain item2item similarity to guide aggregator:
\begin{equation}
% \small
 \footnotesize
	\begin{split}
	\bm{\alpha} = \textsc{Norm}(\{\alpha_1,\alpha_2&,\alpha_3,\dots\}), \ \text{where} \ \alpha_i = (\mathbf{A}\mathbf{B})_{u,v_i}, \\
	\mathbf{h} = \textsc{Weighted}&\textsc{Mean}(\{\mathbf{v}_1,\mathbf{v}_2,\mathbf{v}_3,\dots\}, \bm{\alpha})\mathbf{W}_{\texttt{agg}}, \\
	\end{split}
\label{itemsimilarity}
\end{equation}
where the $\mathbf{A}\mathbf{B} \in \mathbb{R}^{|\mathcal{U}|\times |\mathcal{V}|}$ denote item similarity matrix. 

\subsubsection{Domain-specific \& domain-shared representations}
Former sections has introduced the UniCDR~\cite{unicdr} aggregator to show how the user representations are generated, this section aims to describe how to obtain domain-specific/shared representations.
For simplicity, we assume an overlapped user has its domain $X$ interacted history $\mathcal{H}^X_u=\{v_1^X,v_2^X,v_3^X,\dots\}$ and (if available) all-domain interacted history as $\mathcal{H}^S_u=\{\mathcal{H}^X_u, \mathcal{H}^Y_u$, $\mathcal{H}^Z_u\}$.
Given the in-domain interaction history and mixed interaction history, the UniCDR forms as:
\begin{equation}
% \small
 \footnotesize
	\begin{split}
	\mathbf{h}^X_u = \lambda_A\cdot\mathbf{u}^X + (1-\lambda_A)&\textsc{Aggregator}^X(\mathbf{u}^X, \mathcal{H}^X_u), \\
	\mathbf{h}^S_u = \lambda_A\cdot\mathbf{u}^S + (1-\lambda_A)\Big(&r^X\textsc{Aggregator}^S(\mathbf{u}^S, \mathcal{H}^X_u) \\+ 
 &r^Y\textsc{Aggregator}^S(\mathbf{u}^S, \mathcal{H}^Y_u) \\+ &r^Z\textsc{Aggregator}^S(\mathbf{u}^S, \mathcal{H}^Z_u)\Big), \\
	\text{where} \ r^X = \frac{|\mathcal{H}^X_u|}{|\mathcal{H}^S_u|}, &\ r^Y = \frac{|\mathcal{H}^Y_u|}{|\mathcal{H}^S_u|}, \ r^Z = \frac{|\mathcal{H}^Z_u|}{|\mathcal{H}^S_u|}
 	\end{split}
\label{allaggregator}
\end{equation}
where $h_u^X/h_u^S$ denote the output specific/shared representations, the adaptation factors $r^X, r^Y, r^Z$ adjust the information rate of different domains, the $\lambda_A \in [0, 1]$ is a hype-parameter to keep original signal, and the $\textsc{Aggregator}^\cdot(\cdot)/\textsc{Aggregator}^S(\cdot)$ are domain-specific/shared interaction aggregators 

\subsection{Mask Module}
Up to now, UniCDR~\cite{unicdr} have obtained the user-item aggregated representations $h_u^X/h_u^Y/h_u^Z/h_u^S$, however, it is hard to claim that representation $h_u^S$ is encoded the ``\textbf{domain-invariant}'' information.
Therefore, to guide our shared aggregator to mine valuable information correctly, UniCDR further utilizes mutual formation maximizing intuition~\cite{mine} to max the shared presentation $h_u^S$ with other domains specific representations $h_u^X/h_u^Y/h_u^Z$ correlation.
As a promising way to implement such an idea, UniCDR first mask some interactions to generate multiple shared/specific samples to make them more discriminative, and then identify the related shared-specific representation pairs by a contrastive module to rectify.

\subsubsection{Mechanism details} Actually, UniCDR introduce the two mechanisms of our framework.

\noindent\textbf{(1) Interaction mask}
Specifically, for a user $u$ in domain $X$, UniCDR randomly drop a portion of item interactions in observed history $\mathcal{H}^X_u$, to construct diverse interaction contexts for our method to contrast with:
\begin{equation}
 \footnotesize
	\begin{split}
	\widetilde{\mathcal{H}}_u^X = \textsc{Mask}(\mathcal{H}_u^X, \bm{m}), \ \text{where} \ m_i = \textsc{Bernoulli}(p),
	\end{split}
\label{interactionmask}
\end{equation}
where the $p$ is the probability of each interaction being removed, and $m_i = 1$ is the flag to remove the interaction.
And the perturbations data $\widetilde{\mathcal{H}}_u^X$ will be input to specific $\textsc{Aggregator}^X(\cdot)$ to generate the final output $\widetilde{\mathbf{h}}_u^X$ by Eq.(\ref{allaggregator}).

\noindent\textbf{(2) Domain mask}
Besides, UniCDR further add a mask mechanism aims to introduce cross-domain signal, which remove the in-domain interactions for shared representation and then contrast with in-domain-specific representation as:
\begin{equation}
% \small
 \footnotesize
	\begin{split}
	\widehat{\mathcal{H}}^S_{u,X} = \{\widetilde{\mathcal{H}}^Y_u, \widetilde{\mathcal{H}}^Z_u\}, \ \widehat{\mathcal{H}}^S_{u,Y} = \{\widetilde{\mathcal{H}}^X_u, \widetilde{\mathcal{H}}^Y_u\}, \widehat{\mathcal{H}}^S_{u,Z} = \{\widetilde{\mathcal{H}}^X_u,  \widetilde{\mathcal{H}}^Y_u\}
	\end{split}
\label{domainmask}
\end{equation}
where $\widehat{\mathcal{H}}^S_{u,X}, \widehat{\mathcal{H}}^S_{u,Y}, \widehat{\mathcal{H}}^S_{u,Z}$  guide the shared $\textsc{Aggregator}^S(\cdot)$ to obtain $\widehat{\mathbf{h}}_{u,X}^S, \widehat{\mathbf{h}}_{u,Y}^S, \widehat{\mathbf{h}}_{u,Z}^S$ by Eq.(\ref{allaggregator}), respectively.

\subsection{Contrastive Module}
Based on the masked interaction data, the contrastive module aims to encourage the shared representation encoding the un-biased information by distinguishing the related shared-specific pairs in feature space, e.g., $(\widetilde{\mathbf{h}}^X_u, \widehat{\mathbf{h}}^S_{u,X})$, $(\widetilde{\mathbf{h}}^Y_u, \widehat{\mathbf{h}}^S_{u,Y})$ and $(\widetilde{\mathbf{h}}^Z_u, \widehat{\mathbf{h}}^S_{u,Z})$.
\subsubsection{Hard contrastive objective} 
UniCDR~\cite{unicdr} try to maximize the score between different combinations of the same user shared-specific representation pair and minimize the score between different users' shared-specific pair:
\begin{equation}
% \small
 \scriptsize
	\begin{split}
		\mathcal{L}^X_{\texttt{con}}&=\!\!\!\!\!\sum_{u\in\mathcal{U}^X}\lbrack-\log\textsc{Disc}^X(\widetilde{\mathbf{h}}^X_{u},\widehat{\mathbf{h}}^S_{u,X}) - \log(1 - \textsc{Disc}^X(\widetilde{\mathbf{h}}^X_{\overline{u}},\widehat{\mathbf{h}}^S_{u,X}))\rbrack \\
		&\text{where}\ \textsc{Disc}^X(\widetilde{\mathbf{h}}^X_{u},\widehat{\mathbf{h}}^S_{u,X}) = \textsc{Sigmoid}\big(\widetilde{\mathbf{h}}^X_{u}\mathbf{W}^X_{\texttt{disc}}(\widehat{\mathbf{h}}^S_{u,X})^\top\big),
	\end{split}
\label{contrastive}
\end{equation}
where $\overline{u} \neq u$ is a random other user, $\textsc{Disc}^\cdot(\cdot)$ are discriminator functions with learnable parameter $\mathbf{W}^\cdot_{\texttt{disc}}$.

According to Eq.(\ref{contrastive}), the domain-shared representation is optimized by $\mathcal{L}_{\texttt{con}}^X$, $\mathcal{L}_{\texttt{con}}^Y$ and $\mathcal{L}_{\texttt{con}}^Z$ simultaneously, thus the shared  $\textsc{Aggregator}^S(\cdot)$ tended to encode the balanced domain-invariant information to maximize these classification log-likelihoods.

\subsection{Model Training}
In UniCDR literature~\cite{unicdr}, it focus on public academic datasets, and the training signal is to fit whether the user clicks with a candidate item or not.
Therefore, UniCDR only need to predict the single-type action interactions.
\subsubsection{Single Action Interaction Prediction}
For example, given some positive and negative user-item pairs in domain $X$, UniCDR defines a \textsc{Score}$^\cdot$($\cdot$) function to reconstruct the observed user-item interactions.
\begin{equation}
% \small
\footnotesize
	\begin{split}
	y_{(u,v_i)}^X& = \textsc{Sigmoid}\big(\textsc{Score}^X(\widetilde{\mathbf{h}}^X_{u} , \mathbf{v}_i^X) , \textsc{Score}^X(\widehat{\mathbf{h}}^S_{u,X} , \mathbf{v}_i^X)\big), \\
		&\mathcal{L}_{\texttt{pred}}^X = \sum_{(u, v_i)\in \mathcal{E}^X}\lbrack- \!\log y_{(u,v_i)}^X\! - \! \log(1 - y_{(u,\overline{v}_i)}^X)\rbrack,  
	\end{split}
\label{predictloss}
\end{equation}
where $(u,v_i)$ is the positive observed interaction and $(u,\overline{v}_i)$ is the negative fake interaction, and the \textsc{Score}$(\cdot)$ is replaceable score function, e.g., MLP or inner-dot.

\subsubsection{Loss Function}
In multi-domain Scenario 5, UniCDR optimize the following prediction and contrastive loss:
\begin{equation}
%\small
 \scriptsize
	\begin{split}
	\mathcal{L}	= \lambda(\mathcal{L}^X_\texttt{pred} + \mathcal{L}^Y_\texttt{pred} + \mathcal{L}^Z_\texttt{pred}) + (1-\lambda)(\mathcal{L}_{\texttt{con}}^X + \mathcal{L}_{\texttt{con}}^Y +\mathcal{L}_{\texttt{con}}^Z).
	\end{split}
\label{totalloss}
\end{equation}
where $\lambda\in (0,1)$ is a hyper-parameter.

\begin{figure}[t]
\begin{center}
\includegraphics[width=9cm,height=4cm]{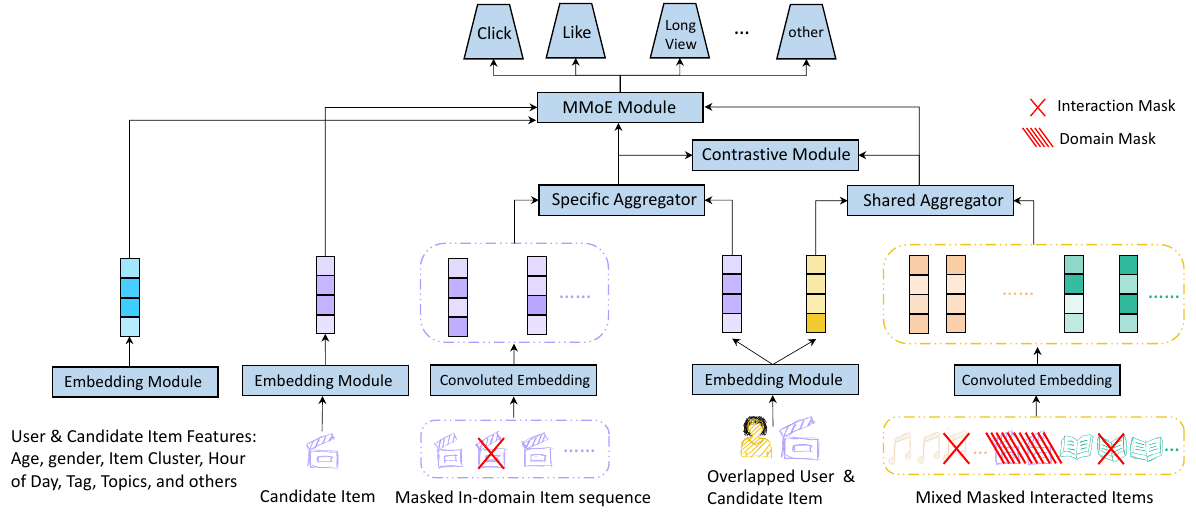}
\caption{The UniCDR+ training procedure in Kuaishou}
\label{framework}
\end{center}
\vspace{-.4cm}
\end{figure}

\subsubsection{Intra-/Inter Domain Recommendation Prediction}
To make downstream task evaluation for a user in domain $X$, we first \textbf{disregard masking mechanisms and input original interaction history} to generate domain-specific and shared representations $\mathbf{h}_u^X$ and $\mathbf{h}_u^S$.
Then, we can predict the following different tasks:
\begin{enumerate}
	\item Intra-domain item recommendation:	
\begin{equation}
%\small
% \footnotesize
\scriptsize
	\begin{split}
	\mathrm{argmax}\ \textsc{Sigmoid}\big(\textsc{Score}^X(\mathbf{h}^X_{u} , \mathbf{v}_i^X) , \textsc{Score}^X(\mathbf{h}^S_{u} , \mathbf{v}_i^X)\big),
	\end{split}
	\label{intraprediction}
\end{equation}
	\item Inter-domain item recommendation:
\begin{equation}
%\small
\footnotesize
	\begin{split}
	\mathrm{argmax}\ \textsc{Sigmoid}\big(\textsc{Score}^Y(\mathbf{h}^S_{u} , \mathbf{v}_i^Y)\big). \\
	\end{split}
	\label{interprediction}
\end{equation}
\end{enumerate}
Such prediction forms also hold in other domains analogously.

\section{Our Extension: UniCDR+ Workflow}
This section presents our extension version, UniCDR+(as shown in Fig.~\ref{framework}), it extends UniCDR in four different aspects to meet industry RecSys demands:
(1) User/Item feature engineering: to introduce the rich statistical features.
(2) Item-level graph neural network: to consider multi-hop item collaborative signals.
(3) Interaction sequence aggregators: modeling the dynamic short-term user interests.
(4) Soft contrastive component: using soft contrastive pair to enhance model effectiveness.
(5) Multi behavioral prediction: the real-world services model needs to predict different actions at same time.

\subsection{Extension: Embedding Module}

\subsubsection{User/Item Feature Embedding}
\label{industryfeature}
Aside from ID features, real-world RecSys usually employs rich context features $\mathcal{F}_u/\mathcal{F}_v$ from the users/items as a part of model inputs to guide model prediction, and can be mainly categorized into the following aspects:
%
% \begin{enumerate}[leftmargin=*,align=left]
\begin{enumerate}
\item \textbf{Categorical features} describes about that 
“gender”, "is following the author", "item cluster", "is high-active user", "device type", etc. 
Generally, we assign a specific embedding $\mathbf{f}_c$ for each feature.
\item \textbf{Numerical features} identify statistical int/float score about that "hour of day", "last behavior timestamp until now", "wallet coins", "history empirical click ratios", etc.
For them, we always truncate to a specific maximum value, and then design a discretization bucket to them to assign their a bucket ID and embeddings $\mathbf{f}_n$.
\item \textbf{Sequential features} are the most important and realtime features for user portrait, which includes the latest and searched click, long-view, like and other item ID sequences. For them, a common way is utilizing the sequence aggregator to produce their embedding.
\end{enumerate}
In public datasets, we only consider the "latest interacted item" sequence feature while ignoring all Categorical/Numerical features for fair experiments setting.
Note that in our Living-Room service, we devise 50+ interaction sequences, and hand-craft about 400+ Categorical/Numerical features for our model UniCDR+ (for simplicity, we denote them as $\mathbf{f} = \{\mathbf{f}_c, \mathbf{f}_n\}$).

\subsection{Extension: Item-level Graph Neural Network}
In industrial RS, the first step for a recommendation request is to retrieval~\cite{tdm} a small candidate set (about thousands items) from the whole item set, and then filter those item candidates by matching~\cite{ccdr} and rank~\cite{din}.
Such a stage is always implemented by the approximate nearest neighbor algorithm~\cite{ann} or graph neural network~\cite{lightgcn} to model item-item correlation.
In UniCDR+, we also consider multi-hop item collaborative signals to enhance item representations.
Specifically, given item embedding matrix $\mathbf{V}_0 = \mathbf{V}$ and interaction matrix $\mathbf{A}\in \{0,1\}^{|\mathcal{U}|\times |\mathcal{V}|}$, we have:
\begin{equation}
% \small
 \footnotesize
	\begin{split}
	\mathbf{V}_1 = \textsc{Norm}(\mathbf{A}^\top)\textsc{Norm}(\mathbf{A})\mathbf{V}_0,
	\end{split}
\label{item_convolution_1}
\end{equation}
where $\mathbf{V}_1$ is the output, $\cdot^\top$ denotes the matrix transpose operator, $\textsc{Norm}(\cdot)$ means the row-normalization operator.
After stacking $L$ layers, we can obtain a series convolution output $\{\mathbf{V}_0,\dots,\mathbf{V}_L\}$.
To exploit the different layer item convolution outputs, we use a $\textsc{Mean}(\cdot)$ function to average them as item convoluted embedding.
\begin{equation}
% \small
 \footnotesize
	\begin{split} 
	\overline{\mathbf{V}} = \textsc{Mean}(\{\mathbf{V}_0,\dots,\mathbf{V}_L\}) + \mathbf{V},
	\end{split}
\label{item_convolution_2}
\end{equation}
where $\overline{\mathbf{V}}$ is the final output. For each domain, we denote the convoluted embedding as $\mathbf{V}^X$, $\mathbf{V}^Y$ and $\mathbf{V}^Z$ for aggregators.

\subsection{Extension: Sequence Aggregator}

\noindent\textbf{(1) Item sequential aggregator}
In recent years, utilizing the sequential item interaction to describe users' dynamic preferences is one hot topic of RS, especially in the Short-Video and Living-Room recommendation.
Indeed, modeling short-term item co-occurrence possibility is closely related to next-word generation in NLP~\cite{gpt3}:
\begin{equation}
% \small
 \footnotesize
	\begin{split}
	\mathbf{h} = &\textsc{Attention}(\textsc{Concat}(\mathbf{u}, \mathbf{v}_{\texttt{1}}), \{\mathbf{v}_1,\mathbf{v}_2,\mathbf{v}_3,\dots\}),
	\end{split}
\label{unicdr_seq_encoder}
\end{equation}
where the $\mathbf{v}_1$ denote the last viewed item embedding.

\noindent\textbf{(2) Deep interest aggregator}
In addition to the above aggregators are suitable to the retrieval process~\cite{tdm} of industry RecSys, there has a famous deep interest aggregator~\cite{din} for fullrank process of industry RecSys.
Unlike retrieval needs to process all items in platform, fullrank is at a later stage in the recommender chain, and only needs to predict scores for a small set of item candidates (typically 50-150 items), and then recommends the highest scoring item for a user request.
Thanks to the small candidates setting, the fullrank process can use more complex mechanisms to maximize recommendation accuracy by generating user interest representations from the item candidate (i.e., the filtered items after retrieval and matching processes):
\begin{equation}
% \small
 \footnotesize
	\begin{split}
	\mathbf{h} = &\textsc{Attention}(\textsc{Concat}(\mathbf{u}, \mathbf{v}_{\texttt{candidate}}), \{\mathbf{v}_1,\mathbf{v}_2,\mathbf{v}_3,\dots\}),
	\end{split}
\label{deepdin}
\end{equation}

\subsection{Extension: Contrastive Module}
\subsubsection{Soft contrastive objective} Different from the hard objective, we further present a soft objective for UniCDR+, which aims at generating more `soft' candidate samples~\cite{mixup} by mixing multiple samples.
For domain $X$, we first mix domain-specific representations in interpolating proportionally manner, and then match their shared representations to estimate the exact proportion:
\begin{equation}
% \small
 \footnotesize
	\begin{split}
		\lambda_{\text{mix}}& = \textsc{Uniform}(0.25, 0.75), \quad
		\widetilde{\mathbf{h}}^X_{u_{ij}} = \lambda_{\text{mix}} \widetilde{\mathbf{h}}^X_{u_i} + (1 - \lambda_{\text{mix}})\widetilde{\mathbf{h}}^X_{u_j},\\
		&\mathcal{L}^X_{\text{con}} = -\sum_{i=1}^{N}\sum_{j=1}^{N}\Big[\mathbf{P}_{ij}\log \frac{\exp\big(\textsc{Disc}^X(\widetilde{\mathbf{h}}^X_{u_{ij}},\widehat{\mathbf{h}}^S_{u_j,X})\big)}{\sum_{k=1}^{N}\exp\big({\textsc{Disc}^X(\widetilde{\mathbf{h}}^X_{u_{ij}},\widehat{\mathbf{h}}^S_{u_k,X})}\big)}\Big],
	\end{split}
\label{unicdr_mixup_contrastive}
\end{equation}
where $\widetilde{\mathbf{h}}^X_{u_{ij}}$ means the mixed domain-specific representation of $u_i$ and $u_j$, $\widehat{\mathbf{h}}^S_{u_j,X}$ is the domain-shared representation of $u_j$, $\lambda_{\text{mix}}$ is a randomly sampled value in [0.25, 0.75], $\textsc{Uniform}(\cdot)$ is the uniform distribution, and $\mathbf{P}_{i} = [0,\dots,\lambda_{\text{mix}},\dots,1-\lambda_{\text{mix}},...,0]$ is a $B$-dim generated label vector (note that only two positions have score $\mathbf{P}_{ii}=\lambda_{\text{mix}}$, $\mathbf{P}_{ij}=1-\lambda_{\text{mix}}$), where $B$ is sample number per batch.

\subsection{Extension: Model Training and Prediction}

\subsubsection{Multiple Action Interactions Prediction}
For real-world services, a user-item interaction includes multiple different prediction objectives, e.g., click, like, long-view, comment, and so on.
Given the collected labels $\hat{y}^{X}_{\cdot}$, we introduce several Mixture-of-Expert-Based~\cite{mmoe,ple} \textsc{Tower}$^\cdot$($\cdot$)$\in [0,1]$ to learn user interest as:
\begin{equation*}
% \small
% \footnotesize
\scriptsize
% \notsotiny
	\begin{split}
	y&_{(u,v_i)}^{X,\texttt{click}} = \textsc{ClickTower}\big(\textsc{Expert}^X(\widetilde{\mathbf{h}}^X_{u},\mathbf{v}_i^X,\mathbf{f}), \textsc{SharedExpert}^X(\widehat{\mathbf{h}}^S_{u,X},\mathbf{v}_i^X,\mathbf{f})\big), \\
	&y_{(u,v_i)}^{X,\texttt{like}} = \textsc{LikeTower}\big(\textsc{Expert}^X(\widetilde{\mathbf{h}}^X_{u},\mathbf{v}_i^X,\mathbf{f}), \textsc{SharedExpert}^X(\widehat{\mathbf{h}}^S_{u,X},\mathbf{v}_i^X,\mathbf{f})\big), \\
		&\mathcal{L}_{\texttt{pred}}^X = \!\!\!\!\!\!\!\!\sum_{(u, v_i)\in \mathcal{E}^X}\sum_{\texttt{action}}\lbrack- \!\hat{y}^{X}_{\texttt{action}}\log y_{(u,v_i)}^{X,\texttt{action}}\! - \! (1 - \hat{y}^{X}_{\texttt{action}})\log(1 - y_{(u,\overline{v}_i)}^{X,\texttt{action}})\rbrack
	\end{split}
\label{mmoepredictloss}
\end{equation*}
where the $\mathbf{f}$ denotes the user and item Categorical/Numerical features in $\S$\ref{industryfeature}, the $\textsc{Expert}^\cdot(\cdot)$/$\textsc{SharedExpert}^\cdot(\cdot)$ are Gated-MLPs, and the $\textsc{Tower}(\cdot)$ is MLPs.

\section{Experiments}

\noindent\textbf{Metrics}
For public CDR tasks, we follow previous works~\cite{conet,bitg} and apply the \textit{leave-one-out} technique~\cite{sampled} (i.e., ranking 1 ground-truth positive item and 999 random negative items) to estimate prediction accuracy at Scenario 1/2/3/4, and enumerate all items to calculate accuracy at Scenario 5.
According to the top-10 items, two metrics are adopted to evaluate retrieval quality, i.e., NDCG and HR (Hit Ratio).
For our fullrank service, we use the AUC and GAUC~\cite{din} (User Group-Based AUC) as offline evaluation metrics.

\begin{table}
% \footnotesize
\scriptsize
\caption{Statistics of five CDR scenarios datasets.}
\resizebox{8cm}{!}{
\begin{tabular}{c|c|cc|ccc}
\toprule
Scenarios & Datasets  &  $|\mathcal{U}|$  &$|\mathcal{V}|$ &Training &Valid &Test  \\ 
\midrule
% \hline
\multirow{2}{*}{Scenario 1}& Sport\_1  &9,928  &30,796 &92,612 &- &8,326   \\ 
\multirow{2}{*}{}& Cloth\_1  &9,928  &39,008 &87,829 &- &7,540   \\ 
\midrule
% \hline
\multirow{2}{*}{Scenario 2}& Sport\_2  &27,328  &12,655 &163,291 &3,589 &3,546    \\ 
\multirow{2}{*}{}& Cloth\_2  &41,829  &17,943 &187,880 &3,156 &3,085    \\ 
\midrule
% \hline
\multirow{5}{*}{Scenario 3}& M1  &7,109  &2,198 &48,302 &3,534 &3,575  \\ 
\multirow{5}{*}{}& M2  &2,697  &1,357 &19,615 &1,375 &1,322    \\ 
\multirow{5}{*}{}& M3  &3,328  &1,245 &23,367 &1,639 &1,689    \\ 
\multirow{5}{*}{}& M4  &5,482  &2,917 &41,226 &2,706 &2,776    \\ 
\multirow{5}{*}{}& M5  &6,466  &9,762 &77,173 &3,207 &3,259   \\ 
\midrule
% \hline
\multirow{2}{*}{Scenario 4}& Food  &-  &29,207 &\multirow{2}{*}{34,117} &2,722 &2,747   \\ 
\multirow{2}{*}{}& Kitchen  &-  &34,886 &\multirow{2}{*}{} &5,451 &5,659   \\ 
\midrule
% \hline
\multirow{3}{*}{Scenario 5}& D1  &231,444  &2,096 &491,098 &13,435 &13,437     \\ 
\multirow{3}{*}{}& D2  &507,715  &595 &1,068,490 &36,013 &35,985     \\ 
\multirow{3}{*}{}& D3  &773,188  &1,312 &3,785,720 &92,659 &92,672    \\ 
\midrule
% \hline
\multirow{2}{*}{Kuaishou}& Short-Video  &\multirow{2}{*}{400-Million} &Billion &\multirow{2}{*}{$\sim$2024/4/27/19} &- &\multirow{2}{*}{2024/4/27/19$\sim$20}     \\ 
\multirow{2}{*}{}& Living-Room  &\multirow{2}{*}{}  &Million &\multirow{2}{*}{} &- &\multirow{2}{*}{}     \\ 
\bottomrule
\end{tabular}
}
\label{datasets}
\vspace{-.4cm}
\end{table}

\subsection{Implementation Details}
\label{imple}
In main public datasets experimental hyper-parameter setting, we follow original literatures~\cite{unicdr, c2dsr} to build training and evaluation environment for UniCDR+.
Specifically, we set embedding dim $d=128$ for Scenario 1/2/3/5 and $d=256$ for Scenario 4, the GNN depth $L=1$ for Scenario 1/3/4/5 and $L=2$ for Scenario 2, the drop-out rate is 0.3, the size per batch $B$ is 1024 for fast training, the sequence length $M$ is 30, the interaction mask rate $p$ is 0.3, the parameter $\lambda_A$ is 0.5, and the loss trade-off $\lambda$ is 0.3, the Frobenius norm regularizer $\lambda_F$ is tuned in [50, 100] with step 5, the training positive/negative ratio is 1/10, and optimize all parameters by Adam.
For public datasets, we train our model 100 epochs and save the model parameters to evaluate accuracy every 5 epochs.

In framework components selection, we find that different tasks needs different aggregators and contrastive objective:
(1) Scenario 1 sets user-attention-pooling aggregator and soft contrastive objective.
(2) Scenario 2 sets mean-pooling aggregator and tuned by hard/soft contrastive objective.
(3) Scenario 3 sets item-similarity-pooling aggregator and hard contrastive objective.
(4) Scenario 4 sets sequential specific/shared aggregator and soft contrastive objective.
(5) Scenario 5 sets mean-pooling aggregator and soft contrastive objective.
(6) The scenario at Kuaishou sets pre-trained item GNN, deep interest aggregator, and hard contrastive objective.

\begin{table*}[th]
\footnotesize
\centering
\caption{The experimental results of five public academic cross-domain scenario datasets.}
\resizebox{\linewidth}{!}{
\setlength{\tabcolsep}{7pt}{
\begin{tabular}{c|c|cccc|cccccc|c}
\toprule
% \hline
%Datasets & Models &F1 &NDCG@3 &NDCG@5 &NDCG@7  &MRR@3 &MRR@5 &MRR@7   &Recall@3 &Recall@5 &Recall@7 \\ 
%\multirow{2}{*}{Datasets} & \multirow{2}{*}{Metrics} & \multicolumn{4}{c}{Single-Domain Methods} & \multicolumn{5}{c}{Cross-Domain Methods}  & \multicolumn{1}{c}{Ours}\\ \cmidrule(lr){3-6}\cmidrule(l){7-11}\cmidrule(l){12-12} & & R@10 &MRR &R@10 &MRR &R@10 &MRR \\
%\multirow{2}{*}{Scenario 1} & \multirow{2}{*}{Metric@10} & \multicolumn{4}{c|}{Single-Domain Methods} & \multicolumn{6}{c|}{Cross-Domain Methods}  & \multicolumn{1}{c}{Ours} \\ \cline{3-6}\cline{7-13}& & BPRMF &NeuMF &NGCF &LightGCN &CoNet &DDTCDR &PPGN &Bi-TGCF &DisenCDR &UniCDR &UniCDR+\\
 Datasets & Metric@10 & BPRMF &NeuMF &NGCF &LightGCN &CoNet &DDTCDR &PPGN &Bi-TGCF &DisenCDR &UniCDR &UniCDR+\\

\midrule
% \hline
\multirow{2}{*}{Sport\_1} & HR  &10.43     &10.74      &13.13    &13.19    &12.09 &11.86 &15.10  &14.83 &17.55 &\underline{18.37} & \textbf{19.14}\\
\multirow{2}{*}{} & NDCG      &5.41     &5.46      &6.87    &6.94    &6.41 &6.37 &8.03  &7.95  &9.46 & \underline{10.98} & \textbf{12.43}\\ 
\cline{2-13}
\multirow{2}{*}{Cloth\_1} &HR  &11.53     &11.18      &13.22    &13.58    &12.40 &12.54 &14.23 &14.68 &16.31 & \underline{17.85} & \textbf{17.98}\\ 
\multirow{2}{*}{} &NDCG      &6.25     &6.02      &6.97    &7.29     &6.62 &7.13 &7.68 &7.93 &9.03 & \underline{11.20} & \textbf{12.22}\\ 
%\midrule
% \hline

%\multirow{2}{*}{Elec} & HR  &15.71     &16.17      &18.55    &19.17      &17.22 &18.47 &21.68  &22.14 &\textbf{24.57} & 22.92 &\underline{23.24} \\
%\multirow{2}{*}{} & NDCG      &9.19     &9.24      &10.87    &10.28    &9.86 &11.08 &11.63  &12.20 &\textbf{14.51} & 13.83 &\underline{14.03} \\ 
%\cline{2-13}
%\multirow{2}{*}{Phone} &HR  &16.32     &15.84      &22.79    &23.25    &17.66 &17.23 &24.54  &25.71 &\textbf{28.76} &24.72 &\underline{25.91} \\ 
%\multirow{2}{*}{} &NDCG      &8.53     &8.02      &12.38    &12.72     &9.30  &8.58 &13.34  &13.93 &\textbf{16.13} & 13.77 & \underline{14.29}\\ 
%\hline
%\hline
%
%\multirow{2}{*}{Sport} & HR  &9.89     &10.11      &16.06    &16.33    &12.88 &12.14 &18.00 &18.63  &20.17 & 18.22\\
%\multirow{2}{*}{} & NDCG      &5.16     &5.19      &8.53    &9.16     &6.91 &6.47 &10.54 &10.11  &11.80 & 10.56\\ 
%\cline{2-12}
%\multirow{2}{*}{Phone} &HR  &13.60     &14.67      &17.07    &16.47   &16.60 &16.17 &20.40  &21.10  &23.55 & 19.10\\ 
%\multirow{2}{*}{} &NDCG      &7.27     &7.80      &9.22    &8.95   &9.15 &8.98 &11.09  &11.25  &12.97 & 10.28\\ 
\bottomrule
% \hline
%\hline

%\multirow{2}{*}{Elec} & HR  &20.65     &20.08      &20.20    &19.97   &21.09 &21.26 &21.70  &\underline{21.85}  &21.61 &23.71 & \\
%\multirow{2}{*}{} & NDCG      &11.66     &11.79      &11.74    &10.73   &11.89  &12.61 &13.10 &\underline{12.36}  &12.25 &13.56 & \\ 
%\cline{2-13}
%\multirow{2}{*}{Cloth} &HR  &9.47     &10.84      &10.86    &11.24   &10.37  &11.35 &11.47 &12.98  &\underline{13.11} &15.13 & \\ 
%\multirow{2}{*}{} &NDCG      &5.07     &5.80      &5.92    &6.11   &5.63  &6.19 &6.38 &\underline{6.88}  &6.80 &8.37 & \\ 
%\hline
\end{tabular}
}
}
%\begin{center}
%\textbf{Boldface} and \underline{underlined} numbers denote the best and runner-up results of all methods, respectively.
%\end{center}
\label{unicdr_overlap_intra}
\end{table*}

\begin{table*}[th]
\footnotesize
% \scriptsize
\centering
\caption{The experimental results of dual-user-static-inter setting}
\setlength{\tabcolsep}{9.5pt}{
\resizebox{\linewidth}{!}{
\begin{tabular}{c|c|ccc|cccccc|c}
\toprule
% \hline
%Datasets & Models &F1 &NDCG@3 &NDCG@5 &NDCG@7  &MRR@3 &MRR@5 &MRR@7   &Recall@3 &Recall@5 &Recall@7 \\ 
%\multirow{2}{*}{Datasets} & \multirow{2}{*}{Metrics} & \multicolumn{4}{c}{Single-Domain Methods} & \multicolumn{5}{c}{Cross-Domain Methods}  & \multicolumn{1}{c}{Ours}\\ \cmidrule(lr){3-6}\cmidrule(l){7-11}\cmidrule(l){12-12} & & R@10 &MRR &R@10 &MRR &R@10 &MRR \\
% \multirow{2}{*}{Scenario 2} & \multirow{2}{*}{Metric@10} & \multicolumn{3}{c|}{Single-Domain Methods} & \multicolumn{6}{c|}{Cross-Domain Methods}  & \multicolumn{1}{c}{Ours} \\ \cline{3-5}\cline{6-12} & & CML &BPRMF &NGCF  &EMCDR &SSCDR &TMCDR &SA-VAE &CDRIB &UniCDR &UniCDR+\\
 Scenario 2 & Metric@10 & CML &BPRMF &NGCF  &EMCDR &SSCDR &TMCDR &SA-VAE &CDRIB &UniCDR &UniCDR+\\
\midrule
% \hline

\multirow{2}{*}{Sport\_2} & HR  &5.82     &5.75      &7.22       &7.44 &7.27 &7.18  &7.51 &\underline{12.04} & 11.20 &\textbf{12.74} \\
\multirow{2}{*}{} & NDCG      &3.29     &3.16      &3.63      &3.71 &3.75 &3.84  &3.72  &6.22 & \underline{7.04} &\textbf{8.36} \\ 
\cline{2-12}
\multirow{2}{*}{Cloth\_2} &HR  &6.97     &6.75      &7.07     &7.29 &6.12 &8.11 &7.21 &12.19 & \underline{12.48} &\textbf{13.02} \\ 
\multirow{2}{*}{} &NDCG      &3.92     &3.26      &3.48        &4.48 &3.06 &5.05 &4.59 &6.81 & \underline{7.52} &\textbf{9.41} \\ 
%\midrule
% \hline

%\multirow{2}{*}{Game} & HR  &2.82     &3.77      &5.14      &4.63 &3.48  &5.36  &5.84 &8.51 & \underline{8.78} & \textbf{9.02}\\
%\multirow{2}{*}{} & NDCG      &1.44     &1.89      &2.73      &2.24 &1.59 &2.58  &2.78 &4.58 & \underline{4.63} & \textbf{4.84}\\ 
%\cline{2-12}
%\multirow{2}{*}{Video} &HR  &3.07     &4.46      &7.41       &7.94 &5.51 &8.85  &7.46 &\textbf{13.17} & 10.74 & \underline{10.76}\\ 
%\multirow{2}{*}{} &NDCG      &1.30     &2.36      &3.87       &4.29 &2.61 &4.41  &3.71 &\textbf{6.49} & 5.89 & \underline{5.97}\\ 
\bottomrule
% \hline
\end{tabular}
}
}
\label{unicdr_overlap_inter}
% \vspace{-.3cm}
\end{table*}

\begin{table*}[th!]
\footnotesize
\centering
\caption{The experimental results of multi-item-static-intra setting}
\resizebox{\linewidth}{!}{
\setlength{\tabcolsep}{5.5pt}{
\begin{tabular}{c|c|cccc|ccccccc|c}
\toprule
% \hline
%Datasets & Models &F1 &NDCG@3 &NDCG@5 &NDCG@7  &MRR@3 &MRR@5 &MRR@7   &Recall@3 &Recall@5 &Recall@7 \\ 
%\multirow{2}{*}{Datasets} & \multirow{2}{*}{Metrics} & \multicolumn{4}{c}{Single-Domain Methods} & \multicolumn{5}{c}{Cross-Domain Methods}  & \multicolumn{1}{c}{Ours}\\ \cmidrule(lr){3-6}\cmidrule(l){7-11}\cmidrule(l){12-12} & & R@10 &MRR &R@10 &MRR &R@10 &MRR \\
% \multirow{2}{*}{Scenario 3} & \multirow{2}{*}{Metric@10} & \multicolumn{4}{c|}{Single-Domain Methods} & \multicolumn{7}{c|}{Cross-Domain Methods}  & \multicolumn{1}{c}{Ours} \\ \cline{3-6}\cline{7-12}\cline{13-14} & &NeuMF &LightGCN &Random Walk &EASE$^R$ &Cross-Stitch &MMoE &Bi-TGCF &STAR &FOREC &M$^3$Rec &UniCDR &UniCDR+\\
 Scenario 3 & Metric@10 &NeuMF &LightGCN &Random Walk &EASE$^R$ &Cross-Stitch &MMoE &Bi-TGCF &STAR &FOREC &M$^3$Rec &UniCDR &UniCDR+\\
\midrule
% \hline
\multirow{2}{*}{M1} & HR  &62.73     &64.73      &64.66    &70.80   &64.46   &65.73 &66.86 &62.93  &65.06 &\textbf{73.13} &69.08 & \underline{69.25}\\
\multirow{2}{*}{} & NDCG      &46.31     &48.30      &48.05    &54.95   &49.15 &48.98 &50.46 &46.57  &52.05 &55.83 & \underline{59.57} & \textbf{61.45}\\ 
\cline{2-14}
\multirow{2}{*}{M2} &HR  &55.60     &52.13      &50.20    &57.40   &54.06 &56.26 &53.46 &54.89  &58.42 &\textbf{60.86} & 58.01 & \underline{58.77}\\ 
\multirow{2}{*}{} &NDCG      &34.84     &32.70      &31.10    &37.92   &36.65   &38.71  &33.43 &35.25  &40.03 &40.04 & \underline{47.52} & \textbf{48.77}\\ 
\cline{2-14}
\multirow{2}{*}{M3} &HR  &60.40     &56.26      &57.53    &63.60   &59.46 &61.53 &58.73 &60.80  &64.13 &\textbf{66.53} & 64.60 & \underline{65.25}\\ 
\multirow{2}{*}{} &NDCG      &36.57     &34.22      &34.89    &40.13   &39.13   &41.30  &35.77 &37.09  &41.88 &43.35 & \underline{53.24} & \textbf{53.90}\\ 
\cline{2-14}
\multirow{2}{*}{M4} &HR  &40.33     &41.14      &39.20    &45.13   &38.93 &38.60 &42.26 &40.20  &41.60 &\textbf{48.46} & \underline{47.52} & 47.30\\ 
\multirow{2}{*}{} &NDCG      &29.83     &31.03      &30.08    &36.63   &29.16   &30.16  &32.76 &29.84  &33.52 &37.98 & \underline{42.54} & \textbf{42.91}\\ 
\cline{2-14}
\multirow{2}{*}{M5} &HR  &12.26     &17.13      &17.73    &19.13   &17.06 &16.66 &17.86 &16.60  &17.46 &\textbf{22.66} & \underline{19.78} & 19.22\\ 
\multirow{2}{*}{} &NDCG      &9.01     &13.16      &14.07    &16.93   &12.51   &11.79  &14.42 &12.48  &13.19 &\textbf{18.63} & 17.04 & \underline{17.54}\\ 
\bottomrule
% \hline
\end{tabular}
}
}
\label{unicdr_item_overlp_intra}
% \vspace{-.3cm}
\end{table*}

\begin{table*}[th!]
%\notsotiny
% \scriptsize
\footnotesize
\centering
\caption{The experimental results of dual-user-sequential-intra setting}
\resizebox{\linewidth}{!}{
 \setlength{\tabcolsep}{8pt}{
\begin{tabular}{c|c|cccc|ccccc|c}
\toprule
% \hline
%Datasets & Models &F1 &NDCG@3 &NDCG@5 &NDCG@7  &MRR@3 &MRR@5 &MRR@7   &Recall@3 &Recall@5 &Recall@7 \\ 
%\multirow{2}{*}{Datasets} & \multirow{2}{*}{Metrics} & \multicolumn{4}{c}{Single-Domain Methods} & \multicolumn{5}{c}{Cross-Domain Methods}  & \multicolumn{1}{c}{Ours}\\ \cmidrule(lr){3-6}\cmidrule(l){7-11}\cmidrule(l){12-12} & & R@10 &MRR &R@10 &MRR &R@10 &MRR \\
% \multirow{2}{*}{Scenario 4} & \multirow{2}{*}{Metric@10} & \multicolumn{4}{c|}{Single-Domain Methods} & \multicolumn{5}{c|}{Cross-Domain Methods}  & \multicolumn{1}{c}{Ours} \\ \cline{3-6}\cline{7-12}& & BPRMF &GRU4Rec &SASRec &SR-GNN &NCF-MLP &CoNet &$\pi$-Net &PSJNet &C$^2$DSR &UniCDR+\\
 Scenario 4 & Metric@10 & BPRMF &GRU4Rec &SASRec &SR-GNN &NCF-MLP &CoNet &$\pi$-Net &PSJNet &C$^2$DSR &UniCDR+\\

\midrule
% \hline
\multirow{2}{*}{Food} & HR  &5.95     &9.11      &11.68    &12.27    &6.86 &6.35 &11.75  &12.45 &\underline{14.54} & \textbf{15.52}\\
\multirow{2}{*}{} & NDCG      &4.03     &6.13      &7.79    &8.35    &4.51 &4.14 &8.13  &8.77  &\underline{9.71} & \textbf{10.84}\\ 
\cline{2-12}
\multirow{2}{*}{Kitchen} &HR  &3.43     &5.22      &6.62    &6.84    &3.65 &3.71 &6.67 &7.15 &\underline{8.18} & \textbf{8.63}\\ 
\multirow{2}{*}{} &NDCG      &1.85     &3.10      &3.93    &4.13     &2.03 &2.11 &3.73 &4.32 &\underline{4.94} & \textbf{5.08}\\ 
%\midrule
%% \hline
%\multirow{2}{*}{Movie} & HR  &4.95     &5.40      &5.20    &5.81      &5.30 &5.35 &6.11  &7.53 &\underline{9.55} & \textbf{10.76}\\
%\multirow{2}{*}{} & NDCG      &2.80     &3.73      &3.69    &3.78    &2.96 &3.01 &4.17  &4.76 &\underline{5.76} & \textbf{6.84}\\ 
%\cline{2-12}
%\multirow{2}{*}{Book} &HR  &2.25     &2.37      &2.75    &2.72    &2.18 &2.19 &2.84  &3.28 &\underline{3.75} & \textbf{3.94}\\ 
%\multirow{2}{*}{} &NDCG      &1.17     &1.52      &1.71    &1.66     &1.26  &1.28 &2.03  &2.35 &\underline{2.45} & \textbf{2.81}\\ 
%\midrule
%% \hline
%\multirow{2}{*}{Entertainment} & HR  &66.08     &73.35      &76.92    &77.17    &71.08 &72.68 &75.11 &76.56  &\underline{79.08} & \textbf{79.93}\\
%\multirow{2}{*}{} & NDCG      &50.11     &51.46      &56.10    &56.47     &50.61 &51.63 &57.63 &\underline{60.07}  &59.35 & \textbf{60.70}\\ 
%\cline{2-12}
%\multirow{2}{*}{Education} &HR  &60.26     &73.21      &76.56    &77.47   &66.43 &68.04 &74.65  &75.08  &\underline{76.21} & \textbf{78.39}\\ 
%\multirow{2}{*}{} &NDCG      &49.27     &56.21      &58.64    &59.69   &50.66 &52.22 &59.32  &60.15  &\underline{61.56} & \textbf{63.44}\\ 
\bottomrule
% \hline
\end{tabular}
}
}
%\begin{center}
%\textbf{Boldface} and \underline{underlined} numbers denote the best and runner-up results of all methods, respectively.
%\end{center}
\label{unicdr_seq_intra}
% \vspace{-.3cm}
\end{table*}

\begin{table*}[th!]
\footnotesize
\centering
\caption{The experimental results of multi-user-static-intra setting}
\resizebox{\linewidth}{!}{
 \setlength{\tabcolsep}{6.5pt}{
\begin{tabular}{c|c|cccc|cccccc|c}
\toprule
% \hline
%Datasets & Models &F1 &NDCG@3 &NDCG@5 &NDCG@7  &MRR@3 &MRR@5 &MRR@7   &Recall@3 &Recall@5 &Recall@7 \\ 
%\multirow{2}{*}{Datasets} & \multirow{2}{*}{Metrics} & \multicolumn{4}{c}{Single-Domain Methods} & \multicolumn{5}{c}{Cross-Domain Methods}  & \multicolumn{1}{c}{Ours}\\ \cmidrule(lr){3-6}\cmidrule(l){7-11}\cmidrule(l){12-12} & & R@10 &MRR &R@10 &MRR &R@10 &MRR \\
%\multirow{2}{*}{Scenario 5} & \multirow{2}{*}{Metric@10} & \multicolumn{4}{c|}{Single-Domain Methods} & \multicolumn{6}{c|}{Cross-Domain Methods}  & \multicolumn{1}{c}{Ours} \\ \cline{3-6}\cline{7-13} & & BPRMF &NeuMF &EASE$^R$ &LightGCN &MMoE &CoNet &Bi-TGCF &GA-MTCDR &HeroGraph &UniCDR &UniCDR+\\
 Scenario 5 & Metric@10 & BPRMF &NeuMF &EASE$^R$ &LightGCN &MMoE &CoNet &Bi-TGCF &GA-MTCDR &HeroGraph &UniCDR &UniCDR+\\

\midrule
% \hline
\multirow{2}{*}{D1} & HR  &19.48     &20.57      &9.15    &25.52   &21.22   &20.60 &26.98 &26.13 &29.73 & \underline{32.60} &\textbf{33.11} \\
\multirow{2}{*}{} & NDCG      &7.66     &7.17      &4.04    &10.60   &8.82 &8.46  &10.64  &10.02 &11.74 & \underline{13.56} &\textbf{14.81} \\ 
\cline{2-13}
\multirow{2}{*}{D2} &HR  &50.45     &52.92      &50.07    &56.18   &56.22 &53.53  &60.48  &59.59 &61.49 &\underline{64.37} &\textbf{64.42} \\ 
\multirow{2}{*}{} &NDCG      &33.50     &35.73      &28.53    &37.09   &38.63   &37.66  &47.19 &47.67 &49.57 & \underline{50.48} &\textbf{51.37} \\ 
\cline{2-13}
\multirow{2}{*}{D3} &HR  &64.87     &64.53      &50.40    &67.13   &65.71 &65.90  &72.88  &73.32 &71.77 & \underline{73.89} &\textbf{74.13} \\ 
\multirow{2}{*}{} &NDCG      &47.69     &48.44      &29.02    &40.49   &47.08   &47.51  &54.15  &57.00 &56.81 & \underline{59.15} &\textbf{61.36} \\
% \cline{2-12}
\bottomrule
% \hline
\end{tabular}
}
}
\label{unicdr_user_overlap_intra}
% \vspace{-.3cm}
\end{table*}

\subsection{Datasets and Metrics}
\noindent\textbf{Datasets}
For a fair experimental setting with previous works, we utilize the preprocessed datasets from Amazon, Ant Group datasets from DisenCDR~\cite{disencdr}, CDRIB~\cite{cdrib}, M$^3$Rec~\cite{m3rec}, C$^2$DSR~\cite{c2dsr} and UniCDR~\cite{unicdr}, all the public datasets can be downloaded at here~\footnote{\url{https://github.com/cjx96/UniCDR} and \url{https://github.com/cjx96/C2DSR}}.
Besides, we also conduct offline experiments in Kuaishou Living-Room recommender system, we train our model before 2024/4/27 19:00, and test its offline performance by an hour sampled data during the 2024/4/27 19:00$\sim$2024/4/27 20:00.
These Scenario statistics are presented in Table~\ref{datasets}.

\begin{table}[t]
\footnotesize
\centering
\caption{Offline results at Living-Room recommendation.}
\setlength{\tabcolsep}{6pt}{
% \resizebox{8cm}{!}{
\begin{tabular}{c|c|c|c|c}
\toprule
% \hline
%Datasets & Models &F1 &NDCG@3 &NDCG@5 &NDCG@7  &MRR@3 &MRR@5 &MRR@7   &Recall@3 &Recall@5 &Recall@7 \\ 
%\multirow{2}{*}{Datasets} & \multirow{2}{*}{Metrics} & \multicolumn{4}{c}{Single-Domain Methods} & \multicolumn{5}{c}{Cross-Domain Methods}  & \multicolumn{1}{c}{Ours}\\ \cmidrule(lr){3-6}\cmidrule(l){7-11}\cmidrule(l){12-12} & & R@10 &MRR &R@10 &MRR &R@10 &MRR \\

% \multirow{2}{*}{Methods} & \multirow{2}{*}{Metrics@10} & \multicolumn{2}{c}{Food} &\multicolumn{2}{c}{Kitchen}\\ \cline{3-6}&  &Static &Seq &Static &Seq \\
% \multirow{2}{*}{Actions} & \multirow{2}{*}{Metrics} & 
Actions &Metrics &DLRM(Base) &DLRM(UniCDR+) &\#Improve \\

% \hline
\hline
% \midrule
\multirow{2}{*}{Click}& AUC  &81.81 &82.20 & +0.38\\
\multirow{2}{*}{} & GAUC     &63.18 &63.92 & +0.73\\ 

% \midrule
\hline
\multirow{2}{*}{Long-View}& AUC  &84.62 &85.16 & +0.54 \\
\multirow{2}{*}{} & GAUC    &68.45 &70.12 & +1.67 \\ 

% \midrule
\hline
\multirow{2}{*}{Like}& AUC  &90.78 &91.33 & +0.55 \\
\multirow{2}{*}{} & GAUC    &72.56 &73.78 & +1.22 \\ 

% \midrule
\hline
\multirow{2}{*}{Comment}& AUC  &93.00 &93.39 & +0.38\\
\multirow{2}{*}{} & GAUC     &71.92 &73.19 & +1.26 \\ 

\bottomrule
\end{tabular}
}

\label{kuaishouresult}
\vspace{-.4cm}
\end{table}

\subsection{Public Datasets Performance}
Table~\ref{unicdr_overlap_intra} report the results about single-domain (first group) and cross-domain (second group) approaches on 5 public CDR Scenarios, and we have following observations:
(1) Comparing with single-domain methods, cross-domain methods generally outperform the corresponding group methods, which indicates that designing elaborate transfer policies is essential.
(2) These cross-domain methods have evolved from coarse-grained implicit knowledge transferring to fine-grained explicit knowledge transferring to reach better results (e.g., CoNet~\cite{conet}$\Rightarrow$Bi-TGCF~\cite{bitg}$\Rightarrow$DisenCDR~\cite{disencdr}, EMCDR~\cite{emcdr}$\Rightarrow$SA-VAE~\cite{savae}$\Rightarrow$CDRIB~\cite{cdrib}, etc.).
(3) Compared with UniCDR, we can see that UniCDR+ shows strong performance in various tasks, and the improvement is statistically in most metrics, which demonstrates our UniCDR+ effectiveness.
(4) Compared with task-expertise methods such as DisenCDR, CDRIB, M$^3$Rec~\cite{m3rec} and C$^2$DSR~\cite{c2dsr}, our UniCDR+ outperform them and shows unique robustness in different Scenarios.

\subsection{Industrial Offline\&Online Performance}
Table.\ref{kuaishouresult} shows the offline performance of UniCDR+ at the Kuaishou Living-Room recommendation services since our online base Deep-Learning-based Recommender Model (DLRM) is a huge model, thus we denoted the UniCDR+ merged variant as DLRM(UniCDR+).
Specifically, our online service needs to tackle billions of user request per day, and the improvement of 0.10\% in offline evaluation AUC and GAUC is significant enough to bring online gains.
According to it, our UniCDR+ could further enhance our online DLRM(base) to predict every action precisely by considering users' short-video interactions, here we show the Click, Long-View, Like, and Comment offline results of Living-Room services.
For the online A/B test, we have replaced the DLRM(base) with DLRM(UniCDR+) for a half month at fullrank process to observe its performances: +0.6\% Click the Living-Room, +0.7\% Long-View, +1.1\% Follow the author, and +1.0\% Living-Room Watch-Time and so on.

\begin{table}[t]
\footnotesize
\centering
\caption{Performance across downstream tasks.}
\setlength{\tabcolsep}{9pt}{
% \resizebox{8cm}{!}{
\begin{tabular}{c|c|cc|cc}
\toprule
% \hline
%Datasets & Models &F1 &NDCG@3 &NDCG@5 &NDCG@7  &MRR@3 &MRR@5 &MRR@7   &Recall@3 &Recall@5 &Recall@7 \\ 
%\multirow{2}{*}{Datasets} & \multirow{2}{*}{Metrics} & \multicolumn{4}{c}{Single-Domain Methods} & \multicolumn{5}{c}{Cross-Domain Methods}  & \multicolumn{1}{c}{Ours}\\ \cmidrule(lr){3-6}\cmidrule(l){7-11}\cmidrule(l){12-12} & & R@10 &MRR &R@10 &MRR &R@10 &MRR \\

\multirow{2}{*}{Methods} & \multirow{2}{*}{Metrics@10} & \multicolumn{2}{c|}{Sport\_2} &\multicolumn{2}{c}{Cloth\_2}\\ \cline{3-6}&  &Intra &Inter &Intra &Inter \\
% \hline
% \hline
\midrule
\multirow{2}{*}{DisenCDR}& HR     &19.39      &$\star$      &19.03    &$\star$ \\
\multirow{2}{*}{} & NDCG        &12.62      &$\star$       &12.05    &$\star$ \\ 
% \hline
\midrule
\multirow{2}{*}{CDRIB}&HR      &$\star$    &10.92 &$\star$ &11.28 \\ 
\multirow{2}{*}{}&NDCG        &$\star$         &5.85 &$\star$ &5.91 \\ 
% \hline
\midrule
\multirow{4}{*}{UniCDR+}&HR      &20.30    &$\star$ &$\star$ &11.62 \\ 
\multirow{4}{*}{}&NDCG        &12.88        &$\star$ &$\star$ &8.70  \\ 
\cline{2-6}
\multirow{4}{*}{}&HR      &$\star$    &10.56 &18.39 &$\star$ \\ 
\multirow{4}{*}{}&NDCG        &$\star$        &7.69 &12.24 &$\star$  \\ 
% \hline
\bottomrule
\end{tabular}
}
\begin{center}
Remark `$\star$' indicates that this domain is not set as corresponding downstream task.
\end{center}
\label{abl_down_task}
\vspace{-.4cm}
\end{table}

\subsection{Analysis Across Downstream Tasks}
To verify the unique cross-domain ability of our UniCDR+, in this section, we further conduct another study to achieve the intra- and inter-recommendation at the same time.
% (As shown in Fig~\ref{abulation_2_fig}).
Specifically, in Scenario 2, we randomly select some interactions from overlapped users in training set to construct an intra-recommendation environment on the Sport\_2\&Cloth\_2 datasets.
Then, we compare our model with two branches SOTA CDR models, DisenCDR and CDRIB, the results are shown in Table~\ref{abl_down_task}.
From it, we can observe:
(1) Since we remove some training data of overlapped users, the inter-domain prediction results show a slight decrease in both CDRIB and UniCDR+, which reveals the amount of overlapped elements information is vital to train a CDR model.
(2) Comparing with DisenCDR and CDRIB, our UniCDR+ shows promising recommendation results in the intra-/inter-recommendation at same time, which further proves the adaptation ability of our UniCDR+ and validates our technique roadmap is suitable for different scenarios.

\subsection{Analysis Across Interaction Types}
This section aims to explore an interesting question: can UniCDR+ transfer knowledge if we know user's static interaction in one domain and sequential interaction in another domain?
% (As shown in Fig~\ref{abulation_1_fig})
To answer this, we conducted an ablation study on Scenario 4.
We first ignore one domain data time-order and then employ the mean-pooling aggregator to support UniCDR+ to generate its representations.
The experimental results are reported in Table ~\ref{interaction_type}. According to it, we can draw the following conclusions:
(1) The sequential setting domain results are much higher than the static setting, which indicates modeling the sequential relationship is vital to describe item sequence data.
(2) We conduct two environment settings, and we can find that our model shows satisfied prediction accuracy in the static domain, which demonstrates our model could transfer knowledge across different interaction types.
It will shed light on interesting research directions to explore more general CDR.

\subsection{Component Effectiveness Discussion}
This section investigates the two model components' effectiveness: the item GNN and soft contrastive objective.
As shown in Table~\ref{component}, for the item aggregator, we conduct 5 variants $L = 1/2/3/4$ to show its robustness on Scenario 1 and 2 (other parameters are listed in Section~\ref{imple}).
According to it, we can find the best results shown in the case $L = 1/2/3$ and show slightly weak performance at case $L = 4$.
The reason might be that the deeper GNN module easily propagates noise interaction information in RS.
Notably, the special case `w/o Item-GNN' denotes that we ignore(without) the module, it shows fiducial performance decreasing than other variants and indicates our item convolution.

For the soft contrastive objective, we give another model variant `w/ Hard' performance in the below of Table~\ref{component}.
Comparing with `w/ Hard', we can find our soft variant `w/ Soft' outperforms than it in many cases,  which demonstrates the soft contrastive objective could further encourage our model to learn better domain-shared user/item representations in Scenario 1/2.
Besides, we find that the hard contrastive objective shows better results in Scenario 3, which indicates that an appropriate contrastive loss is necessary.

\begin{table}[t]
\footnotesize
\centering
\caption{Performance across interaction types.}
\setlength{\tabcolsep}{9pt}{
% \resizebox{8cm}{!}{
\begin{tabular}{c|c|cc|cc}
\toprule
% \hline
%Datasets & Models &F1 &NDCG@3 &NDCG@5 &NDCG@7  &MRR@3 &MRR@5 &MRR@7   &Recall@3 &Recall@5 &Recall@7 \\ 
%\multirow{2}{*}{Datasets} & \multirow{2}{*}{Metrics} & \multicolumn{4}{c}{Single-Domain Methods} & \multicolumn{5}{c}{Cross-Domain Methods}  & \multicolumn{1}{c}{Ours}\\ \cmidrule(lr){3-6}\cmidrule(l){7-11}\cmidrule(l){12-12} & & R@10 &MRR &R@10 &MRR &R@10 &MRR \\

\multirow{2}{*}{Methods} & \multirow{2}{*}{Metrics@10} & \multicolumn{2}{c}{Food} &\multicolumn{2}{c}{Kitchen}\\ \cline{3-6}&  &Static &Seq &Static &Seq \\
% \hline
% \hline
\midrule
\multirow{2}{*}{CoNet}& HR     &6.35      &$\diamond$      &3.71    &$\diamond$ \\
\multirow{2}{*}{} & NDCG        &4.14      &$\diamond$       &2.11    &$\diamond$ \\ 
% \hline
\midrule
\multirow{2}{*}{C$^2$DSR}&HR      &$\diamond$    &14.54 &$\diamond$ &8.18 \\ 
\multirow{2}{*}{}&NDCG        &$\diamond$         &9.71 &$\diamond$ &4.94 \\ 
% \hline
\midrule
\multirow{4}{*}{UniCDR+}&HR      &11.81    &$\diamond$ &$\diamond$ &6.39 \\ 
\multirow{4}{*}{}&NDCG        &7.62        &$\diamond$ &$\diamond$ &3.66  \\ 
\cline{2-6}
\multirow{4}{*}{}&HR      &$\diamond$    &12.43 &5.45 &$\diamond$ \\ 
\multirow{4}{*}{}&NDCG        &$\diamond$        &8.07 &3.06 &$\diamond$  \\ 
% \hline
\bottomrule
\end{tabular}
}
\begin{center}
Remark `$\diamond$' indicates that this domain is not set as the corresponding interaction type.
\end{center}
\label{interaction_type}
\vspace{-.4cm}
\end{table}

\begin{table}[t]
\footnotesize
\centering
\caption{Ablation study of GNN and Contrastive objective.}
\setlength{\tabcolsep}{5pt}{
% \resizebox{8cm}{!}{
\begin{tabular}{c|c|cc|cc}
\hline
%Datasets & Models &F1 &NDCG@3 &NDCG@5 &NDCG@7  &MRR@3 &MRR@5 &MRR@7   &Recall@3 &Recall@5 &Recall@7 \\ 
%\multirow{2}{*}{Datasets} & \multirow{2}{*}{Metrics} & \multicolumn{4}{c}{Single-Domain Methods} & \multicolumn{5}{c}{Cross-Domain Methods}  & \multicolumn{1}{c}{Ours}\\ \cmidrule(lr){3-6}\cmidrule(l){7-11}\cmidrule(l){12-12} & & R@10 &MRR &R@10 &MRR &R@10 &MRR \\

\multirow{2}{*}{Methods} & \multirow{2}{*}{Metrics@10} & \multicolumn{2}{c|}{Scenario 1} &\multicolumn{2}{c}{Scenario 2}\\ \cline{3-6}&  &Sport\_1 &Cloth\_1 &Sport\_2 &Cloth\_2 \\
% Methods & Metrics@10 & Sport\_1 & Cloth\_1 \\
% \hline
\toprule
% \hline
\multirow{2}{*}{w/o Item-GNN}& HR     &18.80      &17.69      &11.31    &12.32 \\
\multirow{2}{*}{} & NDCG        &12.06      &12.05       &8.06    &9.21 \\ 
% \hline
\midrule
\multirow{2}{*}{$L$=1}&HR      &19.14    &17.98 &11.67 &12.61 \\ 
\multirow{2}{*}{}&NDCG        &12.43    &12.22 &8.26 &9.51 \\ 
% \hline
\midrule
\multirow{2}{*}{$L$=2}& HR     &20.51    &18.40 &12.74 &13.02 \\ 
\multirow{2}{*}{} & NDCG        &13.24    &12.86 &8.36 &9.41 \\ 
% \hline
\midrule
\multirow{2}{*}{$L$=3}&HR      &20.25    &18.47 &12.39 &12.55 \\ 
\multirow{2}{*}{}&NDCG        &13.16    &12.83 &8.01 &9.04 \\ 
% \hline
\midrule
\multirow{2}{*}{$L$=4}&HR      &19.86    &18.04 &10.69 &11.40 \\ 
\multirow{2}{*}{}&NDCG        &12.88    &12.38 &7.70 &8.20 \\ 
% \hline
% \hline
\midrule
\midrule
\multirow{2}{*}{w/ Hard}&HR      &18.51    &18.09 &11.42 &12.28 \\ 
\multirow{2}{*}{}&NDCG        &11.40         &11.46 &7.79 &7.98 \\ 
\midrule
\multirow{2}{*}{w/ Soft}&HR      &19.14    &17.98 &12.74 &13.02 \\ 
\multirow{2}{*}{}&NDCG        &12.43         &12.22 &8.36 &9.41 \\ 
% \hline
\bottomrule
\end{tabular}
}
\begin{center}
\end{center}
\label{component}
\vspace{-.4cm}
\end{table}

\section{Related Works}

\label{related_work}
In this section, we review the evolution trajectory of diverse CDRs.

% \subsection{Intra-Cross-Domain Recommendation}
\noindent\textbf{Intra-Cross-Domain Recommendation}
The intra-CDR is proposed to mitigate the data sparsity issue in single-domain RS, by fulfilling the users' preferences with his/her other-domains behaviours.
In earlier times, non-deep-learning intra-CDR methods always relied on cluster idea~\cite{cmf}, first preprocessed the user/item sets by K-means and then insert a cluster compact matrix (called `codebook'~\cite{codebook}) to standard MF paradigm to reconstruct interaction matrix.
In later times, deep-leaning intra-CDR methods reach more flexible and personalized transferring to model user-item interactions.
A classic paradigm that inserts an information-fusion network between different domain-expert decoding towers, such as MLP(e.g., CoNet~\cite{conet}, DDTCDR~\cite{ddtcdr}), GNN (e.g., PPGN~\cite{ppgn}, Bi-TGCF~\cite{bitg}, HeroGraph~\cite{herograph}), Attention (e.g., GA-MTCDR~\cite{gamtcdr}).
Based on above paradigm, other works further introduce some regularizers to learn robust shared/specific representations, such as adversarial-based (e.g., DARec~\cite{darec}), VAEs-based (e.g., DisenCDR~\cite{disencdr}) loss.
To some extent, intra-CDR can be seen as a special case of multi-task learning, thus the Mixture-of-Experts style (e.g., Cross-stitch~\cite{crosss}, MMoE~\cite{mmoe}, STAR~\cite{star}) and pretrain-then-finetune paradigms (e.g., FOREC~\cite{forec}, and M$^3$Rec~\cite{m3rec}) are also can be used to tackle intra-domain CDR.

% \subsection{Inter-Cross-Domain Recommendation}
\noindent\textbf{Inter-Cross-Domain Recommendation}
The inter-CDR is utilized to solve the challenging yet important user cold-start problem of RS, using users' source domain behaviors to recommend target items.
In this branch, a famous backbone paradigm is the pipeline-style Embedding-and-Mapping~\cite{emcdr}: (1) for each domain, pre-train user/item embedding by a single-domain method (e.g., BPRMF~\cite{bprmf}, CML~\cite{cml}, NGCF~\cite{ngcf}, LightGCN~\cite{lightgcn}, etc.), (2) according to overlapped user pair, train a mapping function to align user embedding, (3) based on the mapping function, project other non-overlapped users' source embeddings to target domain.
In recent years, many outstanding works iterate and optimize the second steps, such as introducing the user-GNN (e.g., SSCDR~\cite{sscdr}), meta-network (e.g., TMCDR~\cite{tmcdr}).
Besides, there have some jointly learning paradigm inter-CDR methods with soft user embedding alignment regularizers, such as the Kullback–Leibler divergence (e.g., LinkedVAE~\cite{linkedvae}, SA-VAE~\cite{savae}) and information bottleneck (e.g., CDRIB~\cite{cdrib}).

\noindent\textbf{Cross-Domain Sequential Recommendation}
The CDSR aims to model users' dynamic interests, which extend the sequential recommendation~\cite{srgnn,gru4rec,sasrec} to cross-domain setting.
In sequence modeling, the $\pi$-Net~\cite{pinet}, PSJNet~\cite{psjnet} devise gate filter unit to transfer RNN hidden states synchronously.
Besides, some works further introduce the graph signal to enhance CDSR, such as DA-GCN~\cite{dagcn} employ the GNN on the cross-domain item-item transition graph to enhance item representations.
On this basis, the C$^2$DSR~\cite{c2dsr} considers the whole-graph and within-sequence information simultaneously.

Overall, we can find that most CDR methods can be divided into a specific CDR scenario with different research goals.
In this paper, different from previous works, we made a small step on unified CDR to serve different tasks by one model, even different domains with different interaction types.

\section{Conclusions}

This paper proposes UniCDR+, a unified framework for CDR.
First, we review and explain the CDR concepts and the technique roadmap from a transfer learning perspective.
We then detail the comprehensive workflow and implementation of UniCDR+. By following this workflow and advancing the key components like item-GNN, sequence interest aggregator and soft contrastive objective, UniCDR+ successfully features omnipotence in modeling various characteristics, including static/sequential interactions, multi-hop neighbor signals, and hard/soft representation correlations.
Extensive public datasets experiments demonstrate UniCDR+'s strong performance compared to SOTA baselines on various CDR Scenarios.
Besides, we explore valuable and insightful ablation studies, and may shed light on academic research direction to build more powerful CDR. 
Further, we conduct offline evaluation and online A/B test in large industrial scale RecSys, and successful deployment on Kuaishou, serving hundreds of millions of active users every day.

\balance
\bibliographystyle{ACM-Reference-Format}
\bibliography{sample-base-extend.bib}
\end{document}